\documentclass[aps,prl,twocolumn,footinbib,longbibliography,superscriptaddress]{revtex4-1}
\usepackage{amsmath}
\usepackage{amsfonts}
\usepackage{amssymb}
\usepackage{lmodern,dsfont}
\usepackage{graphicx}
\usepackage[usenames,dvipsnames]{xcolor}
\usepackage{bm}
\usepackage{blkarray}
\usepackage{blindtext}
\usepackage[english=american]{csquotes}

\newcommand{\n}{\nonumber}
\newcommand{\nn}{\nonumber \\}

\newcommand{\grad}{\bm{\nabla}}
\renewcommand{\eqref}[1]{Eq.~(\ref{#1})}

\begin{document}

\author{Andreas Dechant}
\affiliation{WPI-Advanced Institute for Materials Research (WPI-AIMR), Tohoku University, Sendai 980-8577, Japan}
\author{Yohei Sakurai}
\affiliation{WPI-Advanced Institute for Materials Research (WPI-AIMR), Tohoku University, Sendai 980-8577, Japan}

\title{Thermodynamic interpretation of Wasserstein distance}
\date{\today}

\begin{abstract}
We derive a relation between the dissipation in a stochastic dynamics and the Wasserstein distance.
We show that the minimal amount of dissipation required to transform an initial state to a final state during a diffusion process is given by the Wasserstein distance between the two states, divided by the total time of the process.
This relation implies a lower bound on the dissipation for any diffusion process in terms of its initial and final state.
Using a lower bound on the Wasserstein distance, we further show that we can give a lower bound on the dissipation in terms of only the mean and convariance matrix of the initial and final state.
We apply this result to derive the optimal forces that minimize the dissipation for given initial and final mean and covariance.
\end{abstract}

\maketitle

For a physical system that can exchange energy with its environment, any finite-time operation performed on this system is necessarily accompanied by dissipation.
This fact is expressed by the second law of thermodynamics: The change in the system's entropy $\Delta S$ is always larger than the heat $-\Delta Q$ exchanged with the environment divided by the temperature $T$ of the environment.
Another way to state this is that there exists a positive quantity
\begin{align}
\Sigma = \Delta S + \frac{\Delta Q}{T} \geq 0 \label{second-law-0} 
\end{align}
called entropy production, which captures the dissipation due to the irreversibility of the process, and which vanishes only when the dynamics is reversible.
In many cases, the effect of the environment on the system can be effectively described as a combination of dissipative and random forces, rendering the dynamics of the system stochastic \cite{Ris86,Gar02,Cof17}.
In this case, the trajectory of the system during a time interval $[0,\mathcal{T}]$ can only be described in terms of probabilities. 
Irreversibility arises due to the fact that the probability of observing a certain trajectory is different from the probability of observing the time-reversed trajectory \cite{Sei12}.
Within this framework, the entropy production $\Sigma$ can be defined as the Kullback-Leibler divergence between the forward and reverse path probability densities; by virtue of this definition it is positive and vanishes only if the dynamics is reversible and satisfies detailed balance \cite{Sei12}.
In general, $\Sigma = 0$ can only be realized for quasistatic processes, i.~e.~in the limit $\mathcal{T} \rightarrow \infty$.
In practice however, we typically want realize a given process in a given time, and thus have to accept a finite dissipation.

While the entropy production is always positive, different ways of performing the same operation may incur more or less dissipation.
It is thus a natural question whether some finite lower bound on the entropy production for a given operation exists and whether this lower bound can be realized in practice, i.~e.~whether we can find an optimal protocol which realizes the desired operation at minimal dissipation.
Such optimal protocols have been obtained in specific examples \cite{Sch07b,Sch07} and also in the more general context of overdamped Langevin equations \cite{Aur11,Aur12}.
In Refs.~\cite{Aur11,Aur12}, it was noted that problem of finding optimal protocols is closely related to the classical Monge-Kantorovich problem \cite{Mon81,Kan42,Vil08} of optimal transport. 
In its fluid-dynamical formulation by Benamou and Brennier \cite{Ben00}, this problem amounts to finding an optimal velocity field which transports a given initial probability density to a final one.
Interestingly, this formulation does not involve any noise and thus the optimal protocols are those that cause the noisy system to evolve like noiseless optimal transport on average \cite{Aur11}.
While this provides a general framework for determining the optimal protocols, the solution can only be obtained explicitly in specific cases.

The first main result of this Letter is based on the equivalence of the dynamical transport problem \cite{Ben00} with a static formulation in terms of a quantity called the Wasserstein distance \cite{Vas69,Vil08}, which measures the distance between the initial and final state.
We obtain a direct connection between the entropy production in diffusive stochastic dynamics and the $L^2$-Wasserstein distance $\mathcal{W}$ between the initial and final state $p_i$ and $p_f$,
\begin{align}
\Sigma \geq \Sigma^* = \frac{1}{\mu T \mathcal{T}} \mathcal{W}(p_i,p_f)^2 \label{min-dissipation} .
\end{align}
Here $\mu$ is the particle mobility, $T$ the temperature and $\mathcal{T}$ the total duration of the process.
This gives an immediate thermodynamic interpretation of the Wasserstein distance: 
It is equal to the minimal dissipation $\Sigma^*$ over all diffusion processes with initial state $p_i$ and final state $p_f$, where $\Sigma = \Sigma^*$ is obtained for the optimal protocols discussed in Refs.~\cite{Aur11,Aur12}.

Due to the equivalence between the dynamical and static optimal transport problems \cite{Ben00,Vil08}, this is formally equivalent to the lower bound obtained in Ref.~\cite{Aur12}.
However, the identification of the minimal dissipation with the Wasserstein distance allows us to exploit known results for the latter.
In particular, explicit computation of $\Sigma^*$ requires knowledge of the optimal protocols and thus is not feasible in many cases.
By contrast, the Wasserstein distance possesses a lower bound \cite{Gel90}, which together with \eqref{min-dissipation} yields our second main result,
\begin{align}
\Sigma \geq \frac{1}{\mu T \mathcal{T}} \bigg( \Vert &\bm{m}_f - \bm{m}_i \Vert^2 \label{dissipation-bound} \\
&+ \text{tr} \Big( \bm{\Xi}_f + \bm{\Xi}_i - 2 \sqrt{\sqrt{\bm{\Xi}_f} \bm{\Xi}_i \sqrt{\bm{\Xi}_f}} \Big) \bigg) \n ,
\end{align}where $\bm{m}_i$ ($\bm{m}_f$) and $\bm{\Xi}_i$ ($\bm{\Xi}_f$) are the mean and covariance matrix of the initial (final) probability density and tr denotes the trace.
On the one hand, this gives an explicit lower bound on the dissipation in terms of measurable quantities, without the need of computing optimal protocols.
On the other hand, since the right-hand side is precisely the Wasserstein distance between two Gaussian distributions, this implies that, if all we are interested in is the initial and final mean and covariance matrix, the smallest dissipation is obtained when the initial and final state are Gaussian.
Thus, a Gaussian dynamics driven by linear forces realizes minimal dissipation for given mean and covariance matrix.
In this case, we show that the optimal protocols can be obtained in full generality.

Intuitively, the equivalence between Wasserstein distance and dissipation is rooted in the overdamped nature of the dynamics: 
Since any displacement of an overdamped particle generates dissipation, it is natural to expect that the minimal dissipation depends on the spatial distance between the initial and final distribution, which is precisely what is expressed by the Wasserstein distance.
We find that even though this direct connection between displacement and dissipation is lost in for underdamped dynamics, the bound \eqref{min-dissipation} still holds when replacing the initial and final distribution by the marginal distributions of the position,
\begin{align}
\Sigma \geq \frac{1}{\mu T \mathcal{T}} \mathcal{W}(p^x_i,p^x_f)^2 \label{min-dissipation-underdamped} .
\end{align}
This surprising result states that, independent of the velocity distribution, the dissipation in a driven underdamped process is bounded from below by the Wasserstein distance between the initial and final position distribution.
Crucially, in contrast to the overdamped case, we can generally not achieve equality by any protocol, i.~e.~the inequality is strict.

\textit{Fokker-Planck equation and entropy production.}
We first consider a $d$-dimensional overdamped diffusion process $\bm{x}(t) = \lbrace x_1(t), \ldots, x_d(t) \rbrace$, whose probability density $p(\bm{x},t)$ and local mean velocity $\bm{\nu}(\bm{x},t)$ evolve according to the Fokker-Planck equation \cite{Ris86,Spe06}
\begin{subequations}
\begin{align}
\partial_t p(\bm{x},t) &= - \grad \big(\bm{\nu}(\bm{x},t) p(\bm{x},t) \big) \label{continuity} \\
\bm{\nu}(\bm{x},t) &= \mu \big( \bm{f}(\bm{x},t) - T \grad \ln p(\bm{x},t) \big) \label{current}
\end{align} \label{fpe}%
\end{subequations}
on the time interval $[0,\mathcal{T}]$ and with given initial density $p(\bm{x},0)$.
Here $\bm{f}(\bm{x},t)$ is a time-dependent force, $\mu$ is the particle mobility and $T$ is the temperature of the environment.
For the dynamics \eqref{fpe}, the entropy production has an explicit expression in terms of the local mean velocity \cite{Sei12,Sek10,Spi12}
\begin{align}
\Sigma = \int_0^\mathcal{T} dt \ \sigma(t), \quad \sigma(t) = \frac{1}{\mu T} \int d\bm{x} \ \big\Vert \bm{\nu}(\bm{x},t) \big\Vert^2 p(\bm{x},t), \label{entropy-FP}
\end{align}
where $\sigma(t) \geq 0$ is the rate of entropy production.
The entropy production admits a decomposition into the heat $\Delta Q$ exchanged with the environment and the change $\Delta S$ in Shannon entropy \cite{Sei12}.
The positivity of $\Sigma$ yields the second law of thermodynamics \eqref{second-law-0}, which allows interpreting $\Sigma = W^\text{irr}/T$ as the irreversible work which amounts to the excess dissipation during the process.

\textit{Minimal dissipation dynamics and optimal transport.}
In the following, we consider the temperature of the environment and mobility to be constant.
We further fix the initial state $p(\bm{x},0) = p_i(\bm{x})$ and the final state $p(\bm{x},\mathcal{T}) = p_f(\bm{x})$.
Then, the goal is to find the stochastic process whose probability density obeys \eqref{fpe}, which transforms $p_i(\bm{x})$ into $p_f(\bm{x})$ during time $\mathcal{T}$ while minimizing the dissipation $\Sigma$.
Mathematically, we thus have to minimize the target functional $\Sigma$ under the constraints imposed by \eqref{fpe} and the initial and final state.
Since the temperature is fixed, this amounts to finding an optimal choice for the force $\bm{f}(\bm{x},t)$.
Incorporating the constraint \eqref{fpe} via a Lagrange multiplier and solving the corresponding Euler-Lagrange equations, we find the condition for a stationary point of $\Sigma$ \cite{Aur11,Aur12}
\begin{subequations}
\begin{align}
\bm{\nu}(\bm{x},t) &= -\mu T \grad \phi(\bm{x},t) \label{force-gradient} \\
\partial_t \phi(\bm{x},t) &= -\frac{1}{2} \mu T \big\Vert \grad \phi(\bm{x},t) \big\Vert^2. \label{hamilton-jacobi} 
\end{align} \label{stationarity-cond}%
\end{subequations}
This means that the optimal local mean velocity obtained as the gradient of a potential $\phi(\bm{x},t)$, which satisfies the Hamilton-Jacobi equation.
As discussed in Refs.~\cite{Aur11,Aur12}, this is precisely equivalent to the noiseless optimal transport problem investigated by Benamou and Brennier \cite{Ben00}, with the local mean velocity taking the place of the deterministic velocity field $\bm{v}(\bm{x},t)$ in the latter.
The optimal force $\bm{f}^*$ is then obtained from the optimal velocity field and probability density $(\bm{v}^*,p^*)$ as $\bm{f}^*(\bm{x},t) = \bm{v}^*(\bm{x},t)/\mu + T \grad \ln p^*(\bm{x},t)$.
We discuss these points in the Supplemental Material.

A central result of optimal transport theory is that the minimum transport cost is equal to the square of the $L^2$-Wasserstein distance between the initial and final probability density \cite{Ben00,Vil08},
\begin{subequations}
\begin{align}
\mathcal{T} \int_0^\mathcal{T} dt \ \big\Vert \bm{\nu}^*&(\bm{x},t) \big\Vert^2 p^*(\bm{x},t) = \mathcal{W}(p_i,p_f)^2 \\
\mathcal{W}(p_i,p_f)^2 &= \inf_{\Pi} \int d\bm{x} \int d\bm{y} \ \Vert \bm{x} - \bm{y} \Vert^2 \Pi(\bm{x},\bm{y}) \label{wasserstein},
\end{align}
\end{subequations}
where the infimum is taken over all joint probability densities $\Pi(\bm{x},\bm{y})$ with marginals $p_i(\bm{x})$ and $p_f(\bm{y})$.
Comparing this to \eqref{entropy-FP}, we immediately obtain \eqref{min-dissipation}.
Thus, transforming $p_i$ to $p_f$ during time $\mathcal{T}$ via the dynamics \eqref{fpe} requires a minimal amount of dissipation $\Sigma^*$, which is given by the square of the Wasserstein distance between initial and final state, divided by the product of mobility, temperature and time.
We remark that in Ref.~\cite{Jor98}, the solution of the Fokker-Planck equation with a gradient force $\bm{f}(\bm{x}) = -\grad U(\bm{x})$ was characterized as the gradient flow of the free energy with respect to the Wasserstein metric.
Since for a gradient force, the free energy difference $\Delta F$ is related to the entropy production via $\Delta F = - T \Sigma$, this provides a maximum entropy principle for the Fokker-Planck equation.
While Ref.~\cite{Jor98} identifies the increments of entropy production with the Wasserstein distance between the infinitesimally different probability densities $p(\bm{x},t+dt)$ and $p(\bm{x},t)$, \eqref{min-dissipation} shows that this relation extends to finite times for the minimum-dissipation process.
This can be understood from a geometrical point of view by noting that optimal transport yields geodesics in Wasserstein space \cite{Vil08}, i.~e.~the space of probability measures on $\mathbb{R}^N$ with the metric induced by the Wasserstein distance.
While the evolution of the probability density given by \eqref{fpe} describes a curve in the Wasserstein space, the length of this curve corresponds to the distance between the endpoints only if the curve is a geodesic; for general curves, we obtain the inequality \eqref{min-dissipation}.

While the Wasserstein distance is in general difficult to compute explicitly, there is a useful lower bound \cite{Gel90},
\begin{align}
\mathcal{W}(p_i,p_f)^2 \geq \Vert &\bm{m}_f - \bm{m}_i \Vert^2 \label{wasserstein-bound} \\
&+ \text{tr} \Big( \bm{\Xi}_f + \bm{\Xi}_i - 2 \sqrt{\sqrt{\bm{\Xi}_f} \bm{\Xi}_i \sqrt{\bm{\Xi}_f}} \Big) \n ,
\end{align}
where $\bm{m}_i$ ($\bm{m}_f$) and $\bm{\Xi}_i$ ($\bm{\Xi}_f$) are the mean and covariance matrix of the initial (final) probability density and tr denotes the trace.
Note that equality holds if both the initial and final density are Gaussian.
Combining this with \eqref{min-dissipation}, we obtain the lower bound \eqref{dissipation-bound} on dissipation for arbitrary diffusion processes.
Importantly, this lower bound depends only on the means and covariances of the initial and final probability densities, and can thus be directly evaluated in an experimental setting.
Such lower bounds on the dissipation have recently received much attention in the form of thermodynamic uncertainty relations \cite{Bar15,Gin16,Pie17,Hor17,Dec17,Dec18,Pie18}.
These relations may be seen as complementary to \eqref{dissipation-bound}, since they are valid for steady state situations, whereas the latter provides a bound in terms of the change of the system's state. 
While \eqref{dissipation-bound} is written as a lower bound on the dissipation, we can also interpret it as a lower bound on the time required to transform the initial density into the final one at given dissipation, which complements some recently derived speed limits for stochastic dynamics \cite{Oku18,Shi18}.

\textit{Underdamped dynamics.}
While for overdamped dynamics, the entropy production is directly related to the local mean velocity via \eqref{entropy-FP}, the situation is different for underdamped dynamics \cite{Ris86,Gar02,Cof17},
\begin{subequations}
\begin{align}
\partial_t p(\bm{x},\bm{v},t) &= - \grad_x \big(\bm{v} p \big) - \frac{1}{m} \grad_v \big((\bm{f}^\text{rev} + \bm{f}^\text{irr}) p \big) \\
\bm{f}^\text{rev}(\bm{x},\bm{v},t) &=  \bm{f}(\bm{x},t) \\
\bm{f}^\text{irr}(\bm{x},\bm{v},t) &= -\gamma \Big( \bm{v} + \frac{T}{m} \grad_v \ln p(\bm{x},\bm{v},t) \Big) ,
\end{align} \label{kke}%
\end{subequations}
where $m$, $\bm{x}$ and $\bm{v}$ are the mass, position and velocity of the diffusing particle, respectively. 
$\gamma = 1/\mu$ the friction coefficient, $T$ the temperature of the environment and the superscripts rev and irr denote the reversible and irreversible forces, respectively.
Since the velocity is odd under time-reversal, only the irreversible forces contribute to the entropy production \cite{Sek10,Sei12,Spi12},
\begin{align}
\Sigma = \frac{1}{\gamma T} \int_0^\mathcal{T} dt \int d\bm{x} \int d\bm{v} \ \big\Vert \bm{f}^\text{irr}(\bm{x},\bm{v},t) \big\Vert^2 p(\bm{x},\bm{v},t) \label{entropy-underdamped} .
\end{align}
On the other hand, integrating \eqref{kke} over the velocity, we obtain
\begin{align}
\partial_t p^x(\bm{x},t) = -\grad_x \big( \bm{\nu}(\bm{x},t) p^x(\bm{x},t) \big) \label{x-evolution},
\end{align}
where we defined the local mean velocity as
\begin{align}
\bm{\nu}(\bm{x},t) = \int d\bm{v} \ \bm{v} p(\bm{v},t \vert \bm{x}) \label{meanvel-x}.
\end{align}
Here $p^x(\bm{x},t) = \int d\bm{v} p(\bm{x},\bm{v},t)$ is the marginal probability density of the position and $p(\bm{v},t \vert \bm{x}) = p(\bm{x},\bm{v},t)/p^x(\bm{x},t)$ is the velocity probability density conditioned on the position.
Obviously, \eqref{x-evolution} is of the same form as the continuity equation (\ref{continuity}) and thus, by the duality between the dynamical and static formulation of the optimal transport problem, we have
\begin{align}
\Sigma_\text{cg} &= \frac{\gamma}{T} \int_0^\mathcal{T} dt \int d\bm{x} \ \big\Vert \bm{\nu}(\bm{x},t) \big\Vert^2 p^x(\bm{x},t) \label{entropy-cg} \\
& \geq \frac{\gamma}{T \mathcal{T}} \mathcal{W}(p_i^x,p_f^x)^2 \n ,
\end{align}
in analogy to \eqref{min-dissipation}.
The left-hand side may be thought of as a coarse-grained entropy production which is obtained by only observing the dynamics of the position.
The crucial difference to the overdamped case is that the optimal velocity profile $\bm{\nu}^*$ which leads to equality may not be a valid solution of \eqref{kke} and \eqref{meanvel-x}; nevertheless, the inequality holds.
Next we want to show that $\Sigma \geq \Sigma_\text{cg}$, i.~e.~that the coarse-graninig can at most reduce the observed dissipation.
To do so, we start from \eqref{entropy-underdamped} and write $p(\bm{x},\bm{v},t) = p(\bm{v},t \vert \bm{x}) p^x(\bm{x},t)$,
\begin{align}
\Sigma = \frac{\gamma}{T} \int_0^\mathcal{T} dt \int d\bm{x} \int d\bm{v} \ &\bigg\Vert \bm{v} + \frac{T}{m} \grad_v \ln p(\bm{v},t \vert \bm{x}) \bigg\Vert^2 \nn
&\times p(\bm{v},t \vert \bm{x}) p^x(\bm{x},t) .
\end{align}
Since $p(\bm{v},t\vert \bm{x})$ is a probability density with respect to $\bm{v}$, we may apply the Cauchy-Schwarz inequality to the integral over $\bm{v}$,
\begin{align}
\Sigma \geq \ &\frac{\gamma}{T} \int_0^\mathcal{T} dt \int d\bm{x} \ p^x(\bm{x},t) \\ 
& \times \bigg\Vert \int d\bm{v} \ \bigg( \bm{v} + \frac{T}{m} \grad_v \ln p(\bm{v},t \vert \bm{x}) \bigg) p(\bm{v},t \vert \bm{x}) \bigg\Vert^2 \n .
\end{align}
Assuming natural boundary conditions, i.~e.~$p(\bm{v},t \vert \bm{x}) \rightarrow 0$ as $\Vert \bm{v} \Vert \rightarrow \infty$, the second term in the integral over $\bm{v}$ vanishes and we recover the definition of the local mean velocity \eqref{meanvel-x}.
We thus have $\Sigma \geq \Sigma_\text{cg}$.
Together with \eqref{entropy-cg}, we then obtain \eqref{min-dissipation-underdamped}, which is the analog of \eqref{min-dissipation} for the underdamped case.
Using \eqref{wasserstein-bound}, we can further bound this from below by an expression which only involves the mean and covariance of the marginal position probability density, similar to \eqref{dissipation-bound}.
The inequality \eqref{min-dissipation-underdamped} is surprising, since the right hand side contains no information about the distribution of the velocity, i.~e.~the bound holds independent of the initial and final velocity distributions.
Intuitively, this is due to the fact that, since the friction force in \eqref{kke} is linear, any change in the particle position requires a finite average velocity and thus is accompanied by dissipation, much like in the overdamped case.
We stress that, contrary to the overdamped case, where an optimal force which realizes equality in \eqref{min-dissipation} always exists, this is generally not the case in the underdamped case of \eqref{min-dissipation-underdamped}.
Thus, for a given initial and final position probability density the underdamped dynamics generally exhibit a larger dissipation than the overdamped ones.
As discussed in Ref.~\cite{Bau16}, this is due to the fact that we only have limited control over an underdamped dynamics since the relation between velocity and position is dictated by Newton's equations of motion.

\textit{Dissipative particle transport.}
An interesting consequence of \eqref{dissipation-bound} is the following: 
Suppose we are only interested in moving the mean position of a particle system from $\bm{m}_i$ to $\bm{m}_f$ and changing its covariance matrix from $\bm{\Xi}_i$ to $\bm{\Xi}_f$.
Then, \eqref{wasserstein-bound} tells us that the smallest distance between the respective probability densities, and thus the lowest possible dissipation, is obtained when both the initial and final density are Gaussian densities.
As we show in the Supplemental Material, the optimal force realizing equality in \eqref{dissipation-bound} can in this case be obtained explicitly.
It is given by
\begin{align}
\bm{f}^*&(\bm{x},t) = \bm{\Xi}(t)^{-1} \bigg( \frac{\bm{\dot{\Xi}}(t)}{2 \mu} - T \bigg) \big( \bm{x} - \bm{m}(t) \big) + \frac{\bm{\dot{m}}(t)}{\mu}  \nn
&\text{with} \qquad \bm{m}(t) = \bm{m}_f \frac{t}{\mathcal{T}} + \bm{m}_i \Big( 1 - \frac{t}{\mathcal{T}} \Big)  & \nn
&\text{and} \qquad \sqrt{\bm{\Xi}(t)} = \sqrt{\bm{\Xi}_f} \frac{t}{\mathcal{T}} + \sqrt{\bm{\Xi}_i} \Big(1 - \frac{t}{\mathcal{T}} \Big)  \label{optimal-force-gaussian} , &
\end{align}
where $\bm{\dot{m}}$ and $\bm{\dot{\Xi}}$ indicate a time derivative.
The above expression holds under the assumption that the initial and final covariance matrices $\bm{\Xi}_i$ and $\bm{\Xi}_f$ commute; the expression for the general case, while still linear, is significantly more complicated and given in the Supplemental Material.
The corresponding dynamics has a Gaussian probability density whose mean $\bm{m}(t)$ and square root of its covariance matrix $\sqrt{\bm{\sigma}(t)}$ linearly interpolate between the respective initial and final quantities.
This force, which is linear in the position, yields the smallest possible amount of dissipation for any given initial and final mean and covariance matrix.
We remark that, in the one-dimensional case and for constant mean, this force reduces precisely to the optimal protocol derived in Refs.~\cite{Sch07,Sch07b}.
In Ref.~\cite{Sch07}, these optimal protocols were used to characterize the performance of a stochastic heat engine driven by a linear force.
The finding that the linear dynamics minimize dissipation implies that no non-linear version of such a heat engine can outperform the linear one, and the bounds on efficiency of the linear engine derived in Ref.~\cite{Sch07} are in fact general.

\textit{Discussion.}
An important consequence of the one-to-one relation between the minimal dissipation in diffusion processes and the noiseless transport problem is the existence of a unique force which realizes minimal dissipation.
In the case of Gaussian initial and final density, we derived the explicit expression \eqref{optimal-force-gaussian} for the optimal force, which is always linear with respect to $\bm{x}$.
Such linear forces are an ingredient of many minimal models, since they are often the only examples which can be solved explicitly in full generality.
From this point of view, it is reassuring that minimal dissipation always can be realized within a linear model and, in that sense, no excess dissipation is introduced by restricting oneself to linear models.

The Wasserstein distance \eqref{wasserstein} is not only a central concept of optimal transport, but also plays an important role in geometry, where the optimal transport protocols correspond to geodesics with respect to the Wasserstein distance and can be employed in an equivalent definition of curvature, which extends the concept to e.~g.~discrete spaces.
Given that the definition and properties Wasserstein distance readily extend to curved spaces, it is natural to ask what is the corresponding diffusion process whose dissipation reproduces the Wasserstein distance in this case.
We conjecture that this is closely related to diffusion processes with a space-dependent diffusion matrix, which can be identified as metric tensor \cite{Ris86}.
Another natural question is whether the relation between Wasserstein distance and dissipation introduced in this Letter can be extended to Markov jump dynamics on a discrete state space.
While the definition of both dissipation and Wasserstein distance readily generalize to the discrete case, the main difficulty is whether it is possible to find an appropriate distance measure on the discrete space which connects them.

\begin{acknowledgments}
\textbf{Acknowledgments.} This work was supported by the World Premier International Research Center Initiative (WPI), MEXT, Japan and JSPS Grant-in-Aid for Scientific Research on Innovative Areas "Discrete Geometric Analysis for Materials Design": Grant Number 17H06460. 
\end{acknowledgments}

\bibliography{bib}

\begin{thebibliography}{29}%
\makeatletter
\providecommand \@ifxundefined [1]{%
 \@ifx{#1\undefined}
}%
\providecommand \@ifnum [1]{%
 \ifnum #1\expandafter \@firstoftwo
 \else \expandafter \@secondoftwo
 \fi
}%
\providecommand \@ifx [1]{%
 \ifx #1\expandafter \@firstoftwo
 \else \expandafter \@secondoftwo
 \fi
}%
\providecommand \natexlab [1]{#1}%
\providecommand \enquote  [1]{``#1''}%
\providecommand \bibnamefont  [1]{#1}%
\providecommand \bibfnamefont [1]{#1}%
\providecommand \citenamefont [1]{#1}%
\providecommand \href@noop [0]{\@secondoftwo}%
\providecommand \href [0]{\begingroup \@sanitize@url \@href}%
\providecommand \@href[1]{\@@startlink{#1}\@@href}%
\providecommand \@@href[1]{\endgroup#1\@@endlink}%
\providecommand \@sanitize@url [0]{\catcode `\\12\catcode `\$12\catcode
  `\&12\catcode `\#12\catcode `\^12\catcode `\_12\catcode `\%12\relax}%
\providecommand \@@startlink[1]{}%
\providecommand \@@endlink[0]{}%
\providecommand \url  [0]{\begingroup\@sanitize@url \@url }%
\providecommand \@url [1]{\endgroup\@href {#1}{\urlprefix }}%
\providecommand \urlprefix  [0]{URL }%
\providecommand \Eprint [0]{\href }%
\providecommand \doibase [0]{http://dx.doi.org/}%
\providecommand \selectlanguage [0]{\@gobble}%
\providecommand \bibinfo  [0]{\@secondoftwo}%
\providecommand \bibfield  [0]{\@secondoftwo}%
\providecommand \translation [1]{[#1]}%
\providecommand \BibitemOpen [0]{}%
\providecommand \bibitemStop [0]{}%
\providecommand \bibitemNoStop [0]{.\EOS\space}%
\providecommand \EOS [0]{\spacefactor3000\relax}%
\providecommand \BibitemShut  [1]{\csname bibitem#1\endcsname}%
\let\auto@bib@innerbib\@empty
\bibitem [{\citenamefont {Risken}(1986)}]{Ris86}%
  \BibitemOpen
  \bibfield  {author} {\bibinfo {author} {\bibfnamefont {H.}~\bibnamefont
  {Risken}},\ }\href@noop {} {\emph {\bibinfo {title} {The Fokker-Planck
  Equation}}}\ (\bibinfo  {publisher} {Springer Berlin},\ \bibinfo {year}
  {1986})\BibitemShut {NoStop}%
\bibitem [{\citenamefont {Gardiner}(2002)}]{Gar02}%
  \BibitemOpen
  \bibfield  {author} {\bibinfo {author} {\bibfnamefont {C.~W.}\ \bibnamefont
  {Gardiner}},\ }\href@noop {} {\emph {\bibinfo {title} {Handbook of stochastic
  methods: for physics, chemistry and the natural sciences}}}\ (\bibinfo
  {publisher} {Springer},\ \bibinfo {year} {2002})\BibitemShut {NoStop}%
\bibitem [{\citenamefont {Coffey}\ and\ \citenamefont
  {Kalmykov}(2017)}]{Cof17}%
  \BibitemOpen
  \bibfield  {author} {\bibinfo {author} {\bibfnamefont {W.~M.}\ \bibnamefont
  {Coffey}}\ and\ \bibinfo {author} {\bibfnamefont {Y.~P.}\ \bibnamefont
  {Kalmykov}},\ }\href@noop {} {\emph {\bibinfo {title} {The Langevin equation:
  with applications to stochastic problems in physics, chemistry, and
  electrical engineering}}}\ (\bibinfo  {publisher} {World Scientific},\
  \bibinfo {year} {2017})\BibitemShut {NoStop}%
\bibitem [{\citenamefont {Seifert}(2012)}]{Sei12}%
  \BibitemOpen
  \bibfield  {author} {\bibinfo {author} {\bibfnamefont {U.}~\bibnamefont
  {Seifert}},\ }\bibfield  {title} {\enquote {\bibinfo {title} {Stochastic
  thermodynamics, fluctuation theorems and molecular machines},}\ }\href
  {http://stacks.iop.org/0034-4885/75/i=12/a=126001} {\bibfield  {journal}
  {\bibinfo  {journal} {Rep. Prog. Phys.}\ }\textbf {\bibinfo {volume} {75}},\
  \bibinfo {pages} {126001} (\bibinfo {year} {2012})}\BibitemShut {NoStop}%
\bibitem [{\citenamefont {Schmiedl}\ and\ \citenamefont
  {Seifert}(2007)}]{Sch07b}%
  \BibitemOpen
  \bibfield  {author} {\bibinfo {author} {\bibfnamefont {T.}~\bibnamefont
  {Schmiedl}}\ and\ \bibinfo {author} {\bibfnamefont {U.}~\bibnamefont
  {Seifert}},\ }\bibfield  {title} {\enquote {\bibinfo {title} {Optimal
  finite-time processes in stochastic thermodynamics},}\ }\href {\doibase
  10.1103/PhysRevLett.98.108301} {\bibfield  {journal} {\bibinfo  {journal}
  {Phys. Rev. Lett.}\ }\textbf {\bibinfo {volume} {98}},\ \bibinfo {pages}
  {108301} (\bibinfo {year} {2007})}\BibitemShut {NoStop}%
\bibitem [{\citenamefont {Schmiedl}\ and\ \citenamefont
  {Seifert}(2008)}]{Sch07}%
  \BibitemOpen
  \bibfield  {author} {\bibinfo {author} {\bibfnamefont {T.}~\bibnamefont
  {Schmiedl}}\ and\ \bibinfo {author} {\bibfnamefont {U.}~\bibnamefont
  {Seifert}},\ }\bibfield  {title} {\enquote {\bibinfo {title} {Efficiency at
  maximum power: {An} analytically solvable model for stochastic heat
  engines},}\ }\href {http://stacks.iop.org/0295-5075/81/i=2/a=20003}
  {\bibfield  {journal} {\bibinfo  {journal} {EPL (Europhysics Letters)}\
  }\textbf {\bibinfo {volume} {81}},\ \bibinfo {pages} {20003} (\bibinfo {year}
  {2008})}\BibitemShut {NoStop}%
\bibitem [{\citenamefont {Aurell}\ \emph {et~al.}(2011)\citenamefont {Aurell},
  \citenamefont {Mej\'{\i}a-Monasterio},\ and\ \citenamefont
  {Muratore-Ginanneschi}}]{Aur11}%
  \BibitemOpen
  \bibfield  {author} {\bibinfo {author} {\bibfnamefont {E.}~\bibnamefont
  {Aurell}}, \bibinfo {author} {\bibfnamefont {C.}~\bibnamefont
  {Mej\'{\i}a-Monasterio}}, \ and\ \bibinfo {author} {\bibfnamefont
  {P.}~\bibnamefont {Muratore-Ginanneschi}},\ }\bibfield  {title} {\enquote
  {\bibinfo {title} {Optimal protocols and optimal transport in stochastic
  thermodynamics},}\ }\href {\doibase 10.1103/PhysRevLett.106.250601}
  {\bibfield  {journal} {\bibinfo  {journal} {Phys. Rev. Lett.}\ }\textbf
  {\bibinfo {volume} {106}},\ \bibinfo {pages} {250601} (\bibinfo {year}
  {2011})}\BibitemShut {NoStop}%
\bibitem [{\citenamefont {Aurell}\ \emph {et~al.}(2012)\citenamefont {Aurell},
  \citenamefont {Gawȩdzki}, \citenamefont {Mej{\'i}a-Monasterio},
  \citenamefont {Mohayaee},\ and\ \citenamefont
  {Muratore-Ginanneschi}}]{Aur12}%
  \BibitemOpen
  \bibfield  {author} {\bibinfo {author} {\bibfnamefont {E.}~\bibnamefont
  {Aurell}}, \bibinfo {author} {\bibfnamefont {K.}~\bibnamefont {Gawȩdzki}},
  \bibinfo {author} {\bibfnamefont {C.}~\bibnamefont {Mej{\'i}a-Monasterio}},
  \bibinfo {author} {\bibfnamefont {R.}~\bibnamefont {Mohayaee}}, \ and\
  \bibinfo {author} {\bibfnamefont {P.}~\bibnamefont {Muratore-Ginanneschi}},\
  }\bibfield  {title} {\enquote {\bibinfo {title} {Refined second law of
  thermodynamics for fast random processes},}\ }\href {\doibase
  10.1007/s10955-012-0478-x} {\bibfield  {journal} {\bibinfo  {journal} {J.
  Stat. Phys.}\ }\textbf {\bibinfo {volume} {147}},\ \bibinfo {pages}
  {487--505} (\bibinfo {year} {2012})}\BibitemShut {NoStop}%
\bibitem [{\citenamefont {Monge}(1781)}]{Mon81}%
  \BibitemOpen
  \bibfield  {author} {\bibinfo {author} {\bibfnamefont {G.}~\bibnamefont
  {Monge}},\ }\bibfield  {title} {\enquote {\bibinfo {title} {Memoire sur la
  theorie des deblais et des remblais},}\ }\href
  {https://ci.nii.ac.jp/naid/10018386702/en/} {\bibfield  {journal} {\bibinfo
  {journal} {Histoire de l'Academie Royale des Sciences de Paris}\ } (\bibinfo
  {year} {1781})}\BibitemShut {NoStop}%
\bibitem [{\citenamefont {Kantorovitch}(1942)}]{Kan42}%
  \BibitemOpen
  \bibfield  {author} {\bibinfo {author} {\bibfnamefont {L.}~\bibnamefont
  {Kantorovitch}},\ }\bibfield  {title} {\enquote {\bibinfo {title} {On the
  translocation of masses},}\ }\href@noop {} {\bibfield  {journal} {\bibinfo
  {journal} {C. R. (Doklady) Acad. Sci. URSS (N.S.)}\ }\textbf {\bibinfo
  {volume} {37}},\ \bibinfo {pages} {199--201} (\bibinfo {year}
  {1942})}\BibitemShut {NoStop}%
\bibitem [{\citenamefont {Villani}(2008)}]{Vil08}%
  \BibitemOpen
  \bibfield  {author} {\bibinfo {author} {\bibfnamefont {C.}~\bibnamefont
  {Villani}},\ }\href {https://books.google.co.jp/books?id=hV8o5R7\_5tkC}
  {\emph {\bibinfo {title} {Optimal Transport: Old and New}}},\ Grundlehren der
  mathematischen Wissenschaften\ (\bibinfo  {publisher} {Springer Berlin
  Heidelberg},\ \bibinfo {year} {2008})\BibitemShut {NoStop}%
\bibitem [{\citenamefont {Benamou}\ and\ \citenamefont
  {Brenier}(2000)}]{Ben00}%
  \BibitemOpen
  \bibfield  {author} {\bibinfo {author} {\bibfnamefont {J.-D.}\ \bibnamefont
  {Benamou}}\ and\ \bibinfo {author} {\bibfnamefont {Y.}~\bibnamefont
  {Brenier}},\ }\bibfield  {title} {\enquote {\bibinfo {title} {A computational
  fluid mechanics solution to the {Monge-Kantorovich} mass transfer problem},}\
  }\href@noop {} {\bibfield  {journal} {\bibinfo  {journal} {Numer. Math.}\
  }\textbf {\bibinfo {volume} {84}},\ \bibinfo {pages} {375--393} (\bibinfo
  {year} {2000})}\BibitemShut {NoStop}%
\bibitem [{\citenamefont {Vaserstein}(1969)}]{Vas69}%
  \BibitemOpen
  \bibfield  {author} {\bibinfo {author} {\bibfnamefont {L.~N.}\ \bibnamefont
  {Vaserstein}},\ }\bibfield  {title} {\enquote {\bibinfo {title} {Markov
  processes over denumerable products of spaces, describing large systems of
  automata},}\ }\href {http://mi.mathnet.ru/ppi1811} {\bibfield  {journal}
  {\bibinfo  {journal} {Probl. Peredachi Inf.}\ }\textbf {\bibinfo {volume}
  {5}},\ \bibinfo {pages} {250601} (\bibinfo {year} {1969})}\BibitemShut
  {NoStop}%
\bibitem [{\citenamefont {Gelbrich}(1990)}]{Gel90}%
  \BibitemOpen
  \bibfield  {author} {\bibinfo {author} {\bibfnamefont {M.}~\bibnamefont
  {Gelbrich}},\ }\bibfield  {title} {\enquote {\bibinfo {title} {On a formula
  for the {L${}^2$-Wasserstein} metric between measures on {Euclidean} and
  {Hilbert} spaces},}\ }\href@noop {} {\bibfield  {journal} {\bibinfo
  {journal} {Math. Nachr.}\ }\textbf {\bibinfo {volume} {147}},\ \bibinfo
  {pages} {185--203} (\bibinfo {year} {1990})}\BibitemShut {NoStop}%
\bibitem [{\citenamefont {Speck}\ and\ \citenamefont {Seifert}(2006)}]{Spe06}%
  \BibitemOpen
  \bibfield  {author} {\bibinfo {author} {\bibfnamefont {T.}~\bibnamefont
  {Speck}}\ and\ \bibinfo {author} {\bibfnamefont {U.}~\bibnamefont
  {Seifert}},\ }\bibfield  {title} {\enquote {\bibinfo {title} {Restoring a
  fluctuation-dissipation theorem in a nonequilibrium steady state},}\ }\href
  {http://stacks.iop.org/0295-5075/74/i=3/a=391} {\bibfield  {journal}
  {\bibinfo  {journal} {EPL (Europhys. Lett.)}\ }\textbf {\bibinfo {volume}
  {74}},\ \bibinfo {pages} {391} (\bibinfo {year} {2006})}\BibitemShut
  {NoStop}%
\bibitem [{\citenamefont {Sekimoto}(2010)}]{Sek10}%
  \BibitemOpen
  \bibfield  {author} {\bibinfo {author} {\bibfnamefont {K.}~\bibnamefont
  {Sekimoto}},\ }\href {https://books.google.de/books?id=8Fq7BQAAQBAJ} {\emph
  {\bibinfo {title} {Stochastic Energetics}}},\ Lecture Notes in Physics\
  (\bibinfo  {publisher} {Springer Berlin Heidelberg},\ \bibinfo {year}
  {2010})\BibitemShut {NoStop}%
\bibitem [{\citenamefont {Spinney}\ and\ \citenamefont {Ford}(2012)}]{Spi12}%
  \BibitemOpen
  \bibfield  {author} {\bibinfo {author} {\bibfnamefont {R.~E.}\ \bibnamefont
  {Spinney}}\ and\ \bibinfo {author} {\bibfnamefont {I.~J.}\ \bibnamefont
  {Ford}},\ }\bibfield  {title} {\enquote {\bibinfo {title} {Entropy production
  in full phase space for continuous stochastic dynamics},}\ }\href {\doibase
  10.1103/PhysRevE.85.051113} {\bibfield  {journal} {\bibinfo  {journal} {Phys.
  Rev. E}\ }\textbf {\bibinfo {volume} {85}},\ \bibinfo {pages} {051113}
  (\bibinfo {year} {2012})}\BibitemShut {NoStop}%
\bibitem [{\citenamefont {Jordan}\ \emph {et~al.}(1998)\citenamefont {Jordan},
  \citenamefont {Kinderlehrer},\ and\ \citenamefont {Otto}}]{Jor98}%
  \BibitemOpen
  \bibfield  {author} {\bibinfo {author} {\bibfnamefont {R.}~\bibnamefont
  {Jordan}}, \bibinfo {author} {\bibfnamefont {D.}~\bibnamefont
  {Kinderlehrer}}, \ and\ \bibinfo {author} {\bibfnamefont {F.}~\bibnamefont
  {Otto}},\ }\bibfield  {title} {\enquote {\bibinfo {title} {The variational
  formulation of the fokker--planck equation},}\ }\href@noop {} {\bibfield
  {journal} {\bibinfo  {journal} {SIAM J. Math. Anal.}\ }\textbf {\bibinfo
  {volume} {29}},\ \bibinfo {pages} {1--17} (\bibinfo {year}
  {1998})}\BibitemShut {NoStop}%
\bibitem [{\citenamefont {Barato}\ and\ \citenamefont {Seifert}(2015)}]{Bar15}%
  \BibitemOpen
  \bibfield  {author} {\bibinfo {author} {\bibfnamefont {A.~C.}\ \bibnamefont
  {Barato}}\ and\ \bibinfo {author} {\bibfnamefont {U.}~\bibnamefont
  {Seifert}},\ }\bibfield  {title} {\enquote {\bibinfo {title} {Thermodynamic
  uncertainty relation for biomolecular processes},}\ }\href
  {https://journals.aps.org/prl/abstract/10.1103/PhysRevLett.114.158101}
  {\bibfield  {journal} {\bibinfo  {journal} {Phys. Rev. Lett.}\ }\textbf
  {\bibinfo {volume} {114}},\ \bibinfo {pages} {158101} (\bibinfo {year}
  {2015})}\BibitemShut {NoStop}%
\bibitem [{\citenamefont {Gingrich}\ \emph {et~al.}(2016)\citenamefont
  {Gingrich}, \citenamefont {Horowitz}, \citenamefont {Perunov},\ and\
  \citenamefont {England}}]{Gin16}%
  \BibitemOpen
  \bibfield  {author} {\bibinfo {author} {\bibfnamefont {T.~R.}\ \bibnamefont
  {Gingrich}}, \bibinfo {author} {\bibfnamefont {J.~M.}\ \bibnamefont
  {Horowitz}}, \bibinfo {author} {\bibfnamefont {N.}~\bibnamefont {Perunov}}, \
  and\ \bibinfo {author} {\bibfnamefont {J.~L.}\ \bibnamefont {England}},\
  }\bibfield  {title} {\enquote {\bibinfo {title} {Dissipation bounds all
  steady-state current fluctuations},}\ }\href {\doibase
  10.1103/PhysRevLett.116.120601} {\bibfield  {journal} {\bibinfo  {journal}
  {Phys. Rev. Lett.}\ }\textbf {\bibinfo {volume} {116}},\ \bibinfo {pages}
  {120601} (\bibinfo {year} {2016})}\BibitemShut {NoStop}%
\bibitem [{\citenamefont {Pietzonka}\ \emph {et~al.}(2017)\citenamefont
  {Pietzonka}, \citenamefont {Ritort},\ and\ \citenamefont {Seifert}}]{Pie17}%
  \BibitemOpen
  \bibfield  {author} {\bibinfo {author} {\bibfnamefont {P.}~\bibnamefont
  {Pietzonka}}, \bibinfo {author} {\bibfnamefont {F.}~\bibnamefont {Ritort}}, \
  and\ \bibinfo {author} {\bibfnamefont {U.}~\bibnamefont {Seifert}},\
  }\bibfield  {title} {\enquote {\bibinfo {title} {Finite-time generalization
  of the thermodynamic uncertainty relation},}\ }\href {\doibase
  10.1103/PhysRevE.96.012101} {\bibfield  {journal} {\bibinfo  {journal} {Phys.
  Rev. E}\ }\textbf {\bibinfo {volume} {96}},\ \bibinfo {pages} {012101}
  (\bibinfo {year} {2017})}\BibitemShut {NoStop}%
\bibitem [{\citenamefont {Horowitz}\ and\ \citenamefont
  {Gingrich}(2017)}]{Hor17}%
  \BibitemOpen
  \bibfield  {author} {\bibinfo {author} {\bibfnamefont {J.~M.}\ \bibnamefont
  {Horowitz}}\ and\ \bibinfo {author} {\bibfnamefont {T.~R.}\ \bibnamefont
  {Gingrich}},\ }\bibfield  {title} {\enquote {\bibinfo {title} {Proof of the
  finite-time thermodynamic uncertainty relation for steady-state currents},}\
  }\href {\doibase 10.1103/PhysRevE.96.020103} {\bibfield  {journal} {\bibinfo
  {journal} {Phys. Rev. E}\ }\textbf {\bibinfo {volume} {96}},\ \bibinfo
  {pages} {020103} (\bibinfo {year} {2017})}\BibitemShut {NoStop}%
\bibitem [{\citenamefont {Dechant}\ and\ \citenamefont
  {i.~Sasa}(2018)}]{Dec17}%
  \BibitemOpen
  \bibfield  {author} {\bibinfo {author} {\bibfnamefont {A.}~\bibnamefont
  {Dechant}}\ and\ \bibinfo {author} {\bibfnamefont {S.}~\bibnamefont
  {i.~Sasa}},\ }\bibfield  {title} {\enquote {\bibinfo {title} {{Current
  fluctuations and transport efficiency for general Langevin systems}},}\
  }\href {http://stacks.iop.org/1742-5468/2018/i=6/a=063209} {\bibfield
  {journal} {\bibinfo  {journal} {J. Stat. Mech. Theory E.}\ }\textbf {\bibinfo
  {volume} {2018}},\ \bibinfo {pages} {063209} (\bibinfo {year}
  {2018})}\BibitemShut {NoStop}%
\bibitem [{\citenamefont {Dechant}\ and\ \citenamefont {Sasa}(2018)}]{Dec18}%
  \BibitemOpen
  \bibfield  {author} {\bibinfo {author} {\bibfnamefont {A.}~\bibnamefont
  {Dechant}}\ and\ \bibinfo {author} {\bibfnamefont {S.-i.}\ \bibnamefont
  {Sasa}},\ }\bibfield  {title} {\enquote {\bibinfo {title} {Entropic bounds on
  currents in {Langevin} systems},}\ }\href {\doibase
  10.1103/PhysRevE.97.062101} {\bibfield  {journal} {\bibinfo  {journal} {Phys.
  Rev. E}\ }\textbf {\bibinfo {volume} {97}},\ \bibinfo {pages} {062101}
  (\bibinfo {year} {2018})}\BibitemShut {NoStop}%
\bibitem [{\citenamefont {Pietzonka}\ and\ \citenamefont
  {Seifert}(2018)}]{Pie18}%
  \BibitemOpen
  \bibfield  {author} {\bibinfo {author} {\bibfnamefont {P.}~\bibnamefont
  {Pietzonka}}\ and\ \bibinfo {author} {\bibfnamefont {U.}~\bibnamefont
  {Seifert}},\ }\bibfield  {title} {\enquote {\bibinfo {title} {Universal
  trade-off between power, efficiency, and constancy in steady-state heat
  engines},}\ }\href {\doibase 10.1103/PhysRevLett.120.190602} {\bibfield
  {journal} {\bibinfo  {journal} {Phys. Rev. Lett.}\ }\textbf {\bibinfo
  {volume} {120}},\ \bibinfo {pages} {190602} (\bibinfo {year}
  {2018})}\BibitemShut {NoStop}%
\bibitem [{\citenamefont {Okuyama}\ and\ \citenamefont {Ohzeki}(2018)}]{Oku18}%
  \BibitemOpen
  \bibfield  {author} {\bibinfo {author} {\bibfnamefont {M.}~\bibnamefont
  {Okuyama}}\ and\ \bibinfo {author} {\bibfnamefont {M.}~\bibnamefont
  {Ohzeki}},\ }\bibfield  {title} {\enquote {\bibinfo {title} {Quantum speed
  limit is not quantum},}\ }\href {\doibase 10.1103/PhysRevLett.120.070402}
  {\bibfield  {journal} {\bibinfo  {journal} {Phys. Rev. Lett.}\ }\textbf
  {\bibinfo {volume} {120}},\ \bibinfo {pages} {070402} (\bibinfo {year}
  {2018})}\BibitemShut {NoStop}%
\bibitem [{\citenamefont {Shiraishi}\ \emph {et~al.}(2018)\citenamefont
  {Shiraishi}, \citenamefont {Funo},\ and\ \citenamefont {Saito}}]{Shi18}%
  \BibitemOpen
  \bibfield  {author} {\bibinfo {author} {\bibfnamefont {N.}~\bibnamefont
  {Shiraishi}}, \bibinfo {author} {\bibfnamefont {K.}~\bibnamefont {Funo}}, \
  and\ \bibinfo {author} {\bibfnamefont {K.}~\bibnamefont {Saito}},\ }\bibfield
   {title} {\enquote {\bibinfo {title} {Speed limit for classical stochastic
  processes},}\ }\href {\doibase 10.1103/PhysRevLett.121.070601} {\bibfield
  {journal} {\bibinfo  {journal} {Phys. Rev. Lett.}\ }\textbf {\bibinfo
  {volume} {121}},\ \bibinfo {pages} {070601} (\bibinfo {year}
  {2018})}\BibitemShut {NoStop}%
\bibitem [{\citenamefont {Bauer}\ \emph {et~al.}(2016)\citenamefont {Bauer},
  \citenamefont {Brandner},\ and\ \citenamefont {Seifert}}]{Bau16}%
  \BibitemOpen
  \bibfield  {author} {\bibinfo {author} {\bibfnamefont {M.}~\bibnamefont
  {Bauer}}, \bibinfo {author} {\bibfnamefont {K.}~\bibnamefont {Brandner}}, \
  and\ \bibinfo {author} {\bibfnamefont {U.}~\bibnamefont {Seifert}},\
  }\bibfield  {title} {\enquote {\bibinfo {title} {Optimal performance of
  periodically driven, stochastic heat engines under limited control},}\ }\href
  {\doibase 10.1103/PhysRevE.93.042112} {\bibfield  {journal} {\bibinfo
  {journal} {Phys. Rev. E}\ }\textbf {\bibinfo {volume} {93}},\ \bibinfo
  {pages} {042112} (\bibinfo {year} {2016})}\BibitemShut {NoStop}%
\bibitem [{\citenamefont {Knott}\ and\ \citenamefont {Smith}(1984)}]{Kno84}%
  \BibitemOpen
  \bibfield  {author} {\bibinfo {author} {\bibfnamefont {M.}~\bibnamefont
  {Knott}}\ and\ \bibinfo {author} {\bibfnamefont {C.~S.}\ \bibnamefont
  {Smith}},\ }\bibfield  {title} {\enquote {\bibinfo {title} {On the optimal
  mapping of distributions},}\ }\href {\doibase 10.1007/BF00934745} {\bibfield
  {journal} {\bibinfo  {journal} {J. Optimiz. Theory App.}\ }\textbf {\bibinfo
  {volume} {43}},\ \bibinfo {pages} {39--49} (\bibinfo {year}
  {1984})}\BibitemShut {NoStop}%
\end{thebibliography}%

\onecolumngrid

\section{Supplemental Material}

\subsection{Dissipation and optimal transport}
Here, we consider a slightly more general case than in the main text in that we allow the temperature and the mobility to depend on time, while still assuming that they are homogeneous (i.~e.~independent of space) and isotropic (i.~e.~scalar quantities).
The Fokker-Planck equation in $N$ variables for this situation is \cite{Ris86}
\begin{align}
\partial_t p(\bm{x},t) = - \grad \big( \bm{\nu}(\bm{x},t) p(\bm{x},t) \big), \qquad \bm{\nu}(\bm{x},t) = \mu(t) \big( \bm{f}(\bm{x},t) - T(t) \grad \ln p(\bm{x},t) \big) \label{fpe-sm}
\end{align}
and the entropy production is given by \cite{Sek10,Sei12,Spi12}
\begin{align}
\Sigma = \int_0^\mathcal{T} dt \ \frac{1}{\mu(t) T(t)} \int d\bm{x} \ \big\Vert \bm{\nu}(\bm{x},t) \big\Vert^2 p(\bm{x},t) \label{entropy}.
\end{align}
We know want to minimize the entropy over all dynamics obeying \eqref{fpe-sm} while satisfying $p(\bm{x},0) = p_i(\bm{x})$ and $p(\bm{x},\mathcal{T}) = p_f(\bm{x})$.
The constraint provided by \eqref{fpe-sm} can be incorporated via Lagrange multipliers, i.~e.~we compute
\begin{align}
\inf_{\bm{\nu}, \bm{f}, p, \phi, \bm{\psi}} \int_0^\mathcal{T} dt \int d\bm{x} \ \bigg( &\frac{\Vert \bm{\nu}(\bm{x},t) \Vert^2}{\mu(t) T(t)} p(\bm{x},t) - 2\phi(\bm{x},t) \Big( \partial_t p(\bm{x},t) + \grad \big(\bm{\nu}(\bm{x},t) p(\bm{x},t) \big) \Big) \\
& + 2 \bm{\psi}(\bm{x},t) \Big( \bm{\nu}(\bm{x},t) - \mu(t) \big(\bm{f}(\bm{x},t) - T(t) \grad \ln p(\bm{x},t) \big) \Big) \bigg) \n .
\end{align}
The corresponding Euler-Lagrange equations read
\begin{subequations}
\begin{align}
\bm{\nu}: \quad &2 \frac{p(\bm{x},t)}{\mu(t) T(t)} \bm{\nu}(\bm{x},t) - 2 \phi(\bm{x},t) \grad p(\bm{x},t) + 2 \grad \big(\phi(\bm{x},t) p(\bm{x},t) \big) + 2 \bm{\psi}(\bm{x},t) = 0 \\
\bm{f}: \quad &-2 \mu(t) \bm{\psi}(\bm{x},t) = 0 \\
p: \quad & \frac{\Vert \bm{\nu}(\bm{x},t) \Vert^2}{\mu(t) T(t)} + 2 \partial_t \phi(\bm{x},t) + 2 \bm{\nu}(\bm{x},t) \grad \phi(\bm{x},t) + 2 \mu(t) T(t) \bm{\psi}(\bm{x},t) \frac{\grad p(\bm{x},t)}{p(\bm{x},t)^2} + 2 \mu(t) T(t) \grad \Big( \frac{\bm{\psi}(\bm{x},t)}{p(\bm{x},t)} \Big) = 0 \\
\phi: \quad & \partial_t p(\bm{x},t) + \grad \big(\bm{\nu}(\bm{x},t) p(\bm{x},t) \big) = 0 \\
\bm{\psi}: \quad & \bm{\nu}(\bm{x},t) - \mu(t) \big(\bm{f}(\bm{x},t) - T(t) \grad \ln p(\bm{x},t) \big) = 0 .
\end{align}
\end{subequations}
Simplifying this, we obtain
\begin{subequations}
\begin{align}
\bm{\nu}(\bm{x},t) &= - \mu(t) T(t) \grad \phi(\bm{x},t) \\
\partial_t \phi(\bm{x},t) &=- \frac{1}{2} \mu(t) T(t) \big\Vert \grad \phi(\bm{x},t) \big\Vert^2 \\
\partial_t p(\bm{x},t) &= \mu(t) T(t) \grad \big( \grad \phi(\bm{x},t) p(\bm{x},t) \big) \\
\bm{f}(\bm{x},t) &= - T(t) \grad \big( \phi(\bm{x},t) - \ln p(\bm{x},t) \big) .
\end{align}\label{evolution-optimal}%
\end{subequations}
From top to bottom, these equations tell us that the local mean velocity $\bm{\nu}$ is a gradient field of the potential $\phi$, the potential $\phi$ satisfies the Burgers equation, how $p$ evolves in result to the potential and how to obtain the optimal force from the potential $\phi$ and $p$.
In order to facilitate comparison to the optimal transport problem discussed in Ref.~\cite{Ben00}, we introduce the dimensionless time and space coordinates
\begin{align}
\tau(t) = \frac{\int_0^t ds \ \mu(s) T(s)}{\int_0^\mathcal{T} ds \ \mu(s) T(s)}, \qquad \qquad \bm{y}(\bm{x}) = \frac{1}{\sqrt{\int_0^\mathcal{T} ds \ \mu(s) T(s)}} \bm{x} .
\end{align}
While in position space, this corresponds to a uniform rescaling of the coordinates, the time-transformation is non-linear.
However, since we assume $\mu(t) > 0$ and $T(t) > 0$, the time-transformation is invertible.
In terms of the new coordinates, the minimal entropy production and \eqref{evolution-optimal} can be written as
\begin{subequations}
\begin{align}
\Sigma^* &= \int_0^1 dq \int d\bm{z} \ \big\Vert \grad_z \tilde{\phi}(\bm{z},q) \big\Vert^2 \tilde{p}(\bm{z},q) \\
\partial_q \tilde{\phi}(\bm{z},q) &= - \frac{1}{2} \big\Vert \grad_z \tilde{\phi}(\bm{z},q) \big\Vert^2 \\
\partial_q \tilde{p}(\bm{z},q) &= \grad_z \big( \grad_z \tilde{\phi}(\bm{z},q) \tilde{p}(\bm{z},q) \big),
\end{align} \label{bb-evolution}%
\end{subequations}
where we defined $\tilde{\phi}(\bm{z},q) = \phi(\bm{y}^{-1}(\bm{z}),\tau^{-1}(q))$ and $\tilde{p}(\bm{z},q) = (\int_0^\mathcal{T} ds \ \mu(s) T(s))^{N/2} \ p(\bm{y}^{-1}(\bm{z}),\tau^{-1}(q))$.
This is exactly the optimal transport problem of Ref.~\cite{Ben00}, and from the duality between this problem and the Wasserstein distance \cite{Ben00,Vil08} we have
\begin{align}
\Sigma^* = \widetilde{\mathcal{W}}(\tilde{p}(q=0),\tilde{p}(q=1))^2 .
\end{align}
While $q=0$ corresponds to $t=0$ and $q=1$ to $t = \mathcal{T}$, the Wasserstein distance $\widetilde{\mathcal{W}}$ on the right-hand side is the Wasserstein distance with respect to the distance function on the dimensionless space of the coordinate $\bm{y}$.
However, since the latter is related to the dimensionful coordinate $\bm{x}$ by a simple rescaling, we obtain
\begin{align}
\Sigma^* = \frac{1}{\int_0^\mathcal{T} ds \ \mu(s) T(s)} \mathcal{W}(p_i,p_f)^2,
\end{align}
which is the generalization of Eq.~(2) of the main text to time-dependent temperature and mobility and is readily seen to reduce to the former if temperature and mobility are constant.
As an important remark, we note that, since it describes a geodesic in the corresponding Wasserstein space, the solution to \eqref{bb-evolution} is unique \cite{Vil08} and we may thus infer the solution of \eqref{evolution-optimal} from this unique solution.

\subsection{Optimal protocols for Gaussian dynamics}
As discussed in the main text, if the initial and final probability density $p_i$ and $p_f$ are Gaussian with mean $\bm{m}_i$ and $\bm{m}_f$ and covariance matrix $\bm{\Xi}_i$ and $\bm{\Xi}_f$, respectively, then the Wasserstein distance and thus the minimal entropy production is given by
\begin{subequations}
\begin{align}
\mathcal{W}(p_i,p_f)^2 &= \big\Vert \bm{m}_f - \bm{m}_i \big\Vert^2 + \text{tr} \Big( \bm{\Xi}_f + \bm{\Xi}_i - 2 \sqrt{\sqrt{\bm{\Xi}_f} \bm{\Xi}_i \sqrt{\bm{\Xi}_f}} \Big) \label{wasserstein-gaussian} \\ 
\Sigma^* &= \frac{1}{\mu T \mathcal{T}} \bigg( \big\Vert \bm{m}_f - \bm{m}_i \big\Vert^2 + \text{tr} \Big( \bm{\Xi}_f + \bm{\Xi}_i - 2 \sqrt{\sqrt{\bm{\Xi}_f} \bm{\Xi}_i \sqrt{\bm{\Xi}_f}} \Big) \bigg) \label{min-diss-gaussian} ,
\end{align}
\end{subequations}
where, for simplicity, we assume the mobility and temperature to be constant in the following.
As shown in Ref.~\cite{Kno84}, the optimal coupling in the $L^2$-Wasserstein distance
\begin{align}
\mathcal{W}(p_i,p_f)^2 = \inf_\Pi \int d\bm{x} \int d\bm{y} \ \big\Vert \bm{x} - \bm{y} \big\Vert^2 \Pi(\bm{x},\bm{y}) 
\end{align}
is given by the joint density of the stochastic variable $\bm{x}$ distributed according to a Gaussian distribution $\mathcal{N}[\bm{m}_i,\bm{\Xi}_i]$ with mean $\bm{m}_i$ and covariance matrix $\bm{\Xi}_i$ and a stochastic variable $\bm{y}$, which is a linear transform of $\bm{x}$ with
\begin{align}
\bm{y} = \bm{A} \big(\bm{x}-\bm{m}_i \big) + \bm{m}_f,
\end{align}
where the symmetric, positive definite matrix $\bm{A}$ is given by
\begin{align}
\bm{A} = \sqrt{\bm{\Xi}_i^{-1}} \sqrt{\sqrt{\bm{\Xi}_i} \bm{\Xi}_f \sqrt{\bm{\Xi}_i}} \sqrt{\bm{\Xi}_i^{-1}} .
\end{align}
That is, we have the optimal coupling
\begin{align}
\Pi^*(\bm{x},\bm{y}) = \delta\Big(\bm{y} - \bm{A} \big(\bm{x}-\bm{m}_i \big) + \bm{m}_f \Big) \mathcal{N}[\bm{m}_i,\bm{\Xi}_i](\bm{x}) .
\end{align}
In order to find the optimal force $\bm{f}^*(\bm{x},t)$, we first use the fact that the optimal probability density $p^*(\bm{x},t)$ interpolating between $p_i(\bm{x})$ and $p_f(\bm{x})$ is a constant speed geodesic with respect to the Wasserstein distance \cite{Vil08},
\begin{align}
\mathcal{W}(p_i,p^*(s)) = s \mathcal{W}(p_i,p_f) \label{geodesic} ,
\end{align}
where we introduced the rescaled time $s = t/\mathcal{T}$.
Then, the optimal stochastic variable that interpolates between $\bm{x}$ and $\bm{y}$ is just the linear interpolation
\begin{align}
\bm{z}^*(s) = s \bm{y} + (1-s) \bm{x} .
\end{align}
Using explicit computation, it is straightforward to show that $\bm{z}^*(s)$ is distributed according to a Gaussian distribution $\mathcal{N}[\bm{m}^*(s),\bm{\Xi}^*(s)]$ with mean and covariance matrix given by
\begin{subequations}
\begin{align}
\bm{m}^*(s) &= s \bm{m}_i + (1-s) \bm{m}_f \\
\bm{\Xi}^*(s) &=\sqrt{\bm{\Xi}_i^{-1}} \bigg( s \sqrt{\sqrt{\bm{\Xi}_i} \bm{\Xi}_f \sqrt{\bm{\Xi}_i}} + (1-s) \bm{\Xi}_i \bigg)^2  \sqrt{\bm{\Xi}_i^{-1}}.
\end{align} \label{opt-mean-cov}%
\end{subequations}
Plugging this into the expression \eqref{wasserstein-gaussian} for the Wasserstein distance between two Gaussian distributions, we can check that it indeed satisfies the geodesic property \eqref{geodesic}.
All that is left is to find the Fokker-Planck equation that leads to this time evolution.
We make the linear ansatz
\begin{align}
\partial_t p(\bm{x},t) = -\mu \grad \Big( \underbrace{-\bm{K}(t) \bm{x} + \bm{a}(t)}_{\bm{f}(\bm{x},t)} - T \grad \Big) p(\bm{x},t),
\end{align}
where $\bm{K}(t)$ is a symmetric matrix and $\bm{a}(t)$ is a vector.
For a linear force, if the initial probability density is Gaussian, it remains Gaussian for all times. 
The evolution equations for the mean and covariance matrix are found as
\begin{align}
d_t \bm{m}(t) = \mu \Big( - \bm{K}(t) \bm{m}(t) + \bm{a}(t) \Big), \qquad \qquad d_t \bm{\Xi}(t) = \mu \Big( - \big( \bm{K}(t) \bm{\Xi}(t) + \bm{\Xi}(t) \bm{K}(t) \big) + 2 T \Big) .
\end{align}
This allows us to determine $\bm{a}(t)$ and $\bm{K}(t)$ as
\begin{align}
\bm{a}(t) = \frac{1}{\mu} d_t \bm{m}(t) + \bm{K}(t) \bm{m}(t), \qquad \qquad \bm{K}(t) = T \bm{\Xi}^{-1}(t) -\frac{1}{\mu} \int_0^\infty dr \ e^{-r \bm{\Xi}(t)} d_t \bm{\Xi}(t) e^{-r \bm{\Xi}(t)} .
\end{align}
Using this together with \eqref{opt-mean-cov}, we can finally state the optimal force
\begin{subequations}
\begin{align}
\bm{f}^*(\bm{x},t) &= \frac{d_t \bm{m}^*(t)}{\mu} - \bigg( T \bm{\Xi}^*(t)^{-1} -\frac{1}{\mu} \int_0^\infty dr \ e^{-r \bm{\Xi}^*(t)} d_t \bm{\Xi}^*(t) e^{-r \bm{\Xi}^*(t)} \bigg) \big(\bm{x} - \bm{m}^*(t) \big) \quad \text{with} \\
\bm{m}^*(t) &= \frac{t}{\mathcal{T}} \bm{m}_i + \Big(1-\frac{t}{\mathcal{T}}\Big) \bm{m}_f \\
\bm{\Xi}^*(t) &= \sqrt{\bm{\Xi}_i^{-1}} \bigg( \frac{t}{\mathcal{T}} \sqrt{\sqrt{\bm{\Xi}_i} \bm{\Xi}_f \sqrt{\bm{\Xi}_i}} + \Big(1-\frac{t}{\mathcal{T}} \Big) \bm{\Xi}_i \bigg)^2  \sqrt{\bm{\Xi}_i^{-1}} .
\end{align}
\end{subequations}
Note that the above expressions simplify greatly if $\bm{\Xi}_i$ and $\bm{\Xi}_f$ commute, in which case we obtain
\begin{subequations}
\begin{align}
\bm{f}^*(\bm{x},t) &= \frac{d_t \bm{m}^*(t)}{\mu} - \bigg( T -\frac{1}{2\mu} d_t \bm{\Xi}^*(t) \bigg) \bm{\Xi}^*(t)^{-1} \big(\bm{x} - \bm{m}^*(t) \big) \quad \text{with} \\
\bm{m}^*(t) &= \frac{t}{\mathcal{T}} \bm{m}_i + \Big(1-\frac{t}{\mathcal{T}}\Big) \bm{m}_f \\
\bm{\Xi}^*(t) &= \bigg(\frac{t}{\mathcal{T}} \sqrt{\bm{\Xi}_f} + \Big(1-\frac{t}{\mathcal{T}}\Big) \sqrt{\bm{\Xi}_i} \bigg)^2 .
\end{align}
\end{subequations}
In the commuting case, the mean and the square root of the covariance matrix thus interpolate linearly between the initial and final values; this is the result stated in Eq.~(17) of the main text.

\end{document}